\documentclass[twocolumn,letterpaper,aps,prl,superscriptaddress,showpacs,floatfix]{revtex4}

\usepackage{graphicx}	

\usepackage{xspace}	

\newcommand{\sqsn}{\mbox{$\sqrt{s_{_{NN}}}$}\xspace}

\begin{document}


\title{Azimuthal-angle dependence of charged-pion-interferometry 
measurements with respect to 2$^{\rm nd}$- and $3^{\rm rd}$-order 
event planes in Au$+$Au collisions at $\sqrt{s_{_{NN}}}=200$~GeV}

\newcommand{\abilene}{Abilene Christian University, Abilene, Texas 79699, USA}
\newcommand{\augie}{Department of Physics, Augustana College, Sioux Falls, South Dakota 57197, USA}
\newcommand{\banaras}{Department of Physics, Banaras Hindu University, Varanasi 221005, India}
\newcommand{\barc}{Bhabha Atomic Research Centre, Bombay 400 085, India}
\newcommand{\baruch}{Baruch College, City University of New York, New York, New York, 10010 USA}
\newcommand{\bnlcoll}{Collider-Accelerator Department, Brookhaven National Laboratory, Upton, New York 11973-5000, USA}
\newcommand{\bnlphys}{Physics Department, Brookhaven National Laboratory, Upton, New York 11973-5000, USA}
\newcommand{\caucr}{University of California - Riverside, Riverside, California 92521, USA}
\newcommand{\charlesczech}{Charles University, Ovocn\'{y} trh 5, Praha 1, 116 36, Prague, Czech Republic}
\newcommand{\chonbuk}{Chonbuk National University, Jeonju, 561-756, Korea}
\newcommand{\ciae}{Science and Technology on Nuclear Data Laboratory, China Institute of Atomic Energy, Beijing 102413, P.~R.~China}
\newcommand{\cns}{Center for Nuclear Study, Graduate School of Science, University of Tokyo, 7-3-1 Hongo, Bunkyo, Tokyo 113-0033, Japan}
\newcommand{\colorado}{University of Colorado, Boulder, Colorado 80309, USA}
\newcommand{\columbia}{Columbia University, New York, New York 10027 and Nevis Laboratories, Irvington, New York 10533, USA}
\newcommand{\czechtech}{Czech Technical University, Zikova 4, 166 36 Prague 6, Czech Republic}
\newcommand{\dapnia}{Dapnia, CEA Saclay, F-91191, Gif-sur-Yvette, France}
\newcommand{\debrecen}{Debrecen University, H-4010 Debrecen, Egyetem t{\'e}r 1, Hungary}
\newcommand{\elte}{ELTE, E{\"o}tv{\"o}s Lor{\'a}nd University, H - 1117 Budapest, P{\'a}zm{\'a}ny P. s. 1/A, Hungary}
\newcommand{\ewha}{Ewha Womans University, Seoul 120-750, Korea}
\newcommand{\fit}{Florida Institute of Technology, Melbourne, Florida 32901, USA}
\newcommand{\fsu}{Florida State University, Tallahassee, Florida 32306, USA}
\newcommand{\gsu}{Georgia State University, Atlanta, Georgia 30303, USA}
\newcommand{\hiroshima}{Hiroshima University, Kagamiyama, Higashi-Hiroshima 739-8526, Japan}
\newcommand{\ihepprot}{IHEP Protvino, State Research Center of Russian Federation, Institute for High Energy Physics, Protvino, 142281, Russia}
\newcommand{\illuiuc}{University of Illinois at Urbana-Champaign, Urbana, Illinois 61801, USA}
\newcommand{\inrras}{Institute for Nuclear Research of the Russian Academy of Sciences, prospekt 60-letiya Oktyabrya 7a, Moscow 117312, Russia}
\newcommand{\instpasczech}{Institute of Physics, Academy of Sciences of the Czech Republic, Na Slovance 2, 182 21 Prague 8, Czech Republic}
\newcommand{\isu}{Iowa State University, Ames, Iowa 50011, USA}
\newcommand{\jaea}{Advanced Science Research Center, Japan Atomic Energy Agency, 2-4 Shirakata Shirane, Tokai-mura, Naka-gun, Ibaraki-ken 319-1195, Japan}
\newcommand{\jinrdubna}{Joint Institute for Nuclear Research, 141980 Dubna, Moscow Region, Russia}
\newcommand{\jyvaskyla}{Helsinki Institute of Physics and University of Jyv{\"a}skyl{\"a}, P.O.Box 35, FI-40014 Jyv{\"a}skyl{\"a}, Finland}
\newcommand{\kek}{KEK, High Energy Accelerator Research Organization, Tsukuba, Ibaraki 305-0801, Japan}
\newcommand{\korea}{Korea University, Seoul, 136-701, Korea}
\newcommand{\kurchatov}{Russian Research Center ``Kurchatov Institute", Moscow, 123098 Russia}
\newcommand{\kyoto}{Kyoto University, Kyoto 606-8502, Japan}
\newcommand{\labllr}{Laboratoire Leprince-Ringuet, Ecole Polytechnique, CNRS-IN2P3, Route de Saclay, F-91128, Palaiseau, France}
\newcommand{\lahorelums}{Physics Department, Lahore University of Management Sciences, Lahore, Pakistan}
\newcommand{\lawllnl}{Lawrence Livermore National Laboratory, Livermore, California 94550, USA}
\newcommand{\losalamos}{Los Alamos National Laboratory, Los Alamos, New Mexico 87545, USA}
\newcommand{\lpc}{LPC, Universit{\'e} Blaise Pascal, CNRS-IN2P3, Clermont-Fd, 63177 Aubiere Cedex, France}
\newcommand{\lund}{Department of Physics, Lund University, Box 118, SE-221 00 Lund, Sweden}
\newcommand{\maryland}{University of Maryland, College Park, Maryland 20742, USA}
\newcommand{\mass}{Department of Physics, University of Massachusetts, Amherst, Massachusetts 01003-9337, USA }
\newcommand{\michigan}{Department of Physics, University of Michigan, Ann Arbor, Michigan 48109-1040, USA}
\newcommand{\muenster}{Institut fur Kernphysik, University of Muenster, D-48149 Muenster, Germany}
\newcommand{\muhlenberg}{Muhlenberg College, Allentown, Pennsylvania 18104-5586, USA}
\newcommand{\myongji}{Myongji University, Yongin, Kyonggido 449-728, Korea}
\newcommand{\nagasaki}{Nagasaki Institute of Applied Science, Nagasaki-shi, Nagasaki 851-0193, Japan}
\newcommand{\newmex}{University of New Mexico, Albuquerque, New Mexico 87131, USA }
\newcommand{\nmsu}{New Mexico State University, Las Cruces, New Mexico 88003, USA}
\newcommand{\ohio}{Department of Physics and Astronomy, Ohio University, Athens, Ohio 45701, USA}
\newcommand{\ornl}{Oak Ridge National Laboratory, Oak Ridge, Tennessee 37831, USA}
\newcommand{\orsay}{IPN-Orsay, Universite Paris Sud, CNRS-IN2P3, BP1, F-91406, Orsay, France}
\newcommand{\peking}{Peking University, Beijing 100871, P.~R.~China}
\newcommand{\pnpi}{PNPI, Petersburg Nuclear Physics Institute, Gatchina, Leningrad region, 188300, Russia}
\newcommand{\riken}{RIKEN Nishina Center for Accelerator-Based Science, Wako, Saitama 351-0198, Japan}
\newcommand{\rikjrbrc}{RIKEN BNL Research Center, Brookhaven National Laboratory, Upton, New York 11973-5000, USA}
\newcommand{\rikkyo}{Physics Department, Rikkyo University, 3-34-1 Nishi-Ikebukuro, Toshima, Tokyo 171-8501, Japan}
\newcommand{\saispbstu}{Saint Petersburg State Polytechnic University, St. Petersburg, 195251 Russia}
\newcommand{\saopaulo}{Universidade de S{\~a}o Paulo, Instituto de F\'{\i}sica, Caixa Postal 66318, S{\~a}o Paulo CEP05315-970, Brazil}
\newcommand{\seoulnat}{Seoul National University, Seoul, Korea}
\newcommand{\stonybrkc}{Chemistry Department, Stony Brook University, SUNY, Stony Brook, New York 11794-3400, USA}
\newcommand{\stonycrkp}{Department of Physics and Astronomy, Stony Brook University, SUNY, Stony Brook, New York 11794-3400, USA}
\newcommand{\tenn}{University of Tennessee, Knoxville, Tennessee 37996, USA}
\newcommand{\titech}{Department of Physics, Tokyo Institute of Technology, Oh-okayama, Meguro, Tokyo 152-8551, Japan}
\newcommand{\tsukuba}{Institute of Physics, University of Tsukuba, Tsukuba, Ibaraki 305, Japan}
\newcommand{\vandy}{Vanderbilt University, Nashville, Tennessee 37235, USA}
\newcommand{\waseda}{Waseda University, Advanced Research Institute for Science and Engineering, 17 Kikui-cho, Shinjuku-ku, Tokyo 162-0044, Japan}
\newcommand{\weizmann}{Weizmann Institute, Rehovot 76100, Israel}
\newcommand{\wigner}{Institute for Particle and Nuclear Physics, Wigner Research Centre for Physics, Hungarian Academy of Sciences (Wigner RCP, RMKI) H-1525 Budapest 114, POBox 49, Budapest, Hungary}
\newcommand{\yonsei}{Yonsei University, IPAP, Seoul 120-749, Korea}
\affiliation{\abilene}
\affiliation{\augie}
\affiliation{\banaras}
\affiliation{\barc}
\affiliation{\baruch}
\affiliation{\bnlcoll}
\affiliation{\bnlphys}
\affiliation{\caucr}
\affiliation{\charlesczech}
\affiliation{\chonbuk}
\affiliation{\ciae}
\affiliation{\cns}
\affiliation{\colorado}
\affiliation{\columbia}
\affiliation{\czechtech}
\affiliation{\dapnia}
\affiliation{\debrecen}
\affiliation{\elte}
\affiliation{\ewha}
\affiliation{\fit}
\affiliation{\fsu}
\affiliation{\gsu}
\affiliation{\hiroshima}
\affiliation{\ihepprot}
\affiliation{\illuiuc}
\affiliation{\inrras}
\affiliation{\instpasczech}
\affiliation{\isu}
\affiliation{\jaea}
\affiliation{\jinrdubna}
\affiliation{\jyvaskyla}
\affiliation{\kek}
\affiliation{\korea}
\affiliation{\kurchatov}
\affiliation{\kyoto}
\affiliation{\labllr}
\affiliation{\lahorelums}
\affiliation{\lawllnl}
\affiliation{\losalamos}
\affiliation{\lpc}
\affiliation{\lund}
\affiliation{\maryland}
\affiliation{\mass}
\affiliation{\michigan}
\affiliation{\muenster}
\affiliation{\muhlenberg}
\affiliation{\myongji}
\affiliation{\nagasaki}
\affiliation{\newmex}
\affiliation{\nmsu}
\affiliation{\ohio}
\affiliation{\ornl}
\affiliation{\orsay}
\affiliation{\peking}
\affiliation{\pnpi}
\affiliation{\riken}
\affiliation{\rikjrbrc}
\affiliation{\rikkyo}
\affiliation{\saispbstu}
\affiliation{\saopaulo}
\affiliation{\seoulnat}
\affiliation{\stonybrkc}
\affiliation{\stonycrkp}
\affiliation{\tenn}
\affiliation{\titech}
\affiliation{\tsukuba}
\affiliation{\vandy}
\affiliation{\waseda}
\affiliation{\weizmann}
\affiliation{\wigner}
\affiliation{\yonsei}
\author{A.~Adare} \affiliation{\colorado}
\author{S.~Afanasiev} \affiliation{\jinrdubna}
\author{C.~Aidala} \affiliation{\mass} \affiliation{\michigan}
\author{N.N.~Ajitanand} \affiliation{\stonybrkc}
\author{Y.~Akiba} \affiliation{\riken} \affiliation{\rikjrbrc}
\author{H.~Al-Bataineh} \affiliation{\nmsu}
\author{J.~Alexander} \affiliation{\stonybrkc}
\author{K.~Aoki} \affiliation{\kyoto} \affiliation{\riken}
\author{Y.~Aramaki} \affiliation{\cns}
\author{E.T.~Atomssa} \affiliation{\labllr}
% ...
\collaboration{PHENIX Collaboration} \noaffiliation

\date{\today}

\begin{abstract}


Charged-pion-interferometry measurements were made with respect to 
the 2$^{\rm nd}$- and 3$^{\rm rd}$-order event plane for Au$+$Au 
collisions at $\sqrt{s_{_{NN}}}=200$~GeV.  A strong azimuthal-angle 
dependence of the extracted Gaussian-source radii was observed with 
respect to both the 2$^{\rm nd}$- and 3$^{\rm rd}$-order event 
planes.  The results for the 2$^{\rm nd}$-order dependence indicate 
that the initial eccentricity is reduced during the medium 
evolution, but not reversed in the final state, which is consistent 
with previous results.  In contrast, the results for the 
3$^{\rm rd}$-order dependence indicate that the initial triangular 
shape is significantly reduced and potentially reversed by the end 
of the medium evolution, and that the 3$^{\rm rd}$-order 
oscillations are largely dominated by the dynamical effects from 
triangular flow.

\end{abstract}

	
\maketitle



The quark-gluon plasma (QGP), a state of nuclear matter in which 
quarks and gluons are deconfined, is produced in nuclear collisions 
at sufficiently high 
energy~\cite{wh_phenix,wh_star,wh_phobos,wh_brahms}.  Once formed, 
the QGP expands, cools, and then freezes out into a collection of 
final-state particles.  From extensive measurements of 
final-particle momenta and correlations, a detailed space-time 
picture of the evolution of the QGP is 
emerging~\cite{SolvingPuzzle,VHstudy}, but detailed studies of the 
final space-time distribution of hadrons and an understanding of the 
dependence on the initial-collision geometry are needed to complete 
this picture.

Quantum-statistical interferometry of two identical particles, also 
known as Hanbury Brown and Twiss (HBT) 
interferometry~\cite{hbt_org,GGLP}, provides information on the 
space-time extent of the particle-emitting source.  In heavy-ion 
collisions, hadron interferometry is sensitive to the space-time 
extent of the hadronic system at the time of the last scattering, 
referred to as kinetic freeze-out.  In noncentral collisions of like 
nuclei, the initial density distribution is predominantly elliptical 
in shape, with additional fluctuations~\cite{flow_review}.  There is 
a larger pressure gradient along the minor axis (in plane) of the 
ellipse, compared to that along the major axis (out of plane), and 
this leads to a stronger expansion of the source within the in-plane 
direction.  This phenomenon, elliptic flow, reduces the eccentricity 
of the spatial distribution in the transverse plane, and may even 
reverse the major and minor axes of the initial distributions.  
Previous results are consistent with the picture that the final 
distribution still retains the initial elliptical orientation, 
although with a smaller eccentricity upon freeze-out~\cite{star_azhbt}.

The full set of anisotropic moments of the flow is characterized by 
the Fourier coefficients of the azimuthal distribution of emitted 
particles:  $dN/d\phi \propto 1 + 2\sum 
v_{n}\cos[n(\phi-\Psi_{n})]$, where $\phi$ is the azimuthal angle of 
the particle, $v_{n}$ is the strength of n$^{\rm th}$-order flow 
harmonic, and $\Psi_{n}$ is the $n^{\rm th}$-order event plane, 
where $\Psi_{2}$ and $\Psi_{3}$ are independent~\cite{phenix_vn}.  
Elliptic flow is defined by the 2$^{\rm nd}$-order coefficient 
($n$\,=\,2), but triangular ($n$\,=\,3), quadrangular ($n$\,=\,4), 
and higher-order moments are also present and have been measured in 
both the spatial and momentum distributions in heavy ion 
collisions~\cite{phenix_vn,alice_vn,atlas_vn}.  While the 
higher-order even moments are needed to accurately describe the 
original elliptic shape, the odd moments arise solely through 
fluctuations in the initial spatial distribution.  Depending on 
strength of the fluctuations, flow profile, expansion time, and 
shear viscosity, these initial spatial fluctuations may be preserved 
until freeze-out~\cite{Sergei_qm11,glauber_en}.

In relativistic heavy ion collisions, HBT interferometry with 
respect to different order event planes uniquely probes the 
magnitude of the initial-state fluctuations and the subsequent 
space-time evolution, thereby providing important constraints on the 
dynamics of the QGP.  Here, we present results of azimuthal HBT 
measurements of charged pions with respect to 2$^{\rm nd}$-order 
event plane, as well as the first results with respect to the 
3$^{\rm rd}$-order event plane in Au$+$Au collisions at 
\sqsn~=~200~GeV at central rapidity.  The centrality and transverse 
momentum dependence are also presented.


This analysis is based on data collected in 2007 with the PHENIX 
detector~\cite{PHENIX_overview}.  Collision centrality was 
determined using the measured charge distribution in the beam-beam 
counters (3.0\,$<$\,$|\eta|$\,$<$\,3.9)~\cite{PHENIX_inner}.  The event 
planes, $\Psi_{n}$, were determined using the reaction plane detector 
(RXNP) covering forward and backward angles 
1.0\,$<$\,$|\eta|$\,$<$\,2.8~\cite{RXNP}. The event plane resolution 
${\rm Res}(\Psi_{n})$ was estimated by the two-subevent 
method~\cite{TwoSub} using the $\Psi_{n}$ correlation between the 
RXNP at forward and backward angles, where ${\rm Res}(\Psi_{n})$ is 
defined as $\langle\cos[n(\Psi_{n}-\Psi_{n,{\rm real}})]\rangle$.  
Track and momentum reconstruction of charged particles was performed 
by combining hits from the drift chamber and pad chambers in the 
central spectrometers ($|\eta|$\,$<$\,0.35), where the momentum 
resolution is $\delta p/p \approx 1.3\% \oplus 1.2\% \times 
p$~\cite{momres}.  Charged pions were identified by combining 
time-of-flight from the electromagnetic 
calorimeters~\cite{PHENIX_EMC} covering azimuthal angle 
$\Delta\phi$\,$=$\,$\pi/2$, with reconstructed momentum and 
trajectory in the magnetic field.  Particles within two standard 
deviations of the peak of charged pions in mass-squared 
distributions were identified as pions up to a momentum of 
$\sim$1~GeV/$c$.


\begin{figure*}[tb]
\begin{center}
\includegraphics[width=0.99\linewidth]{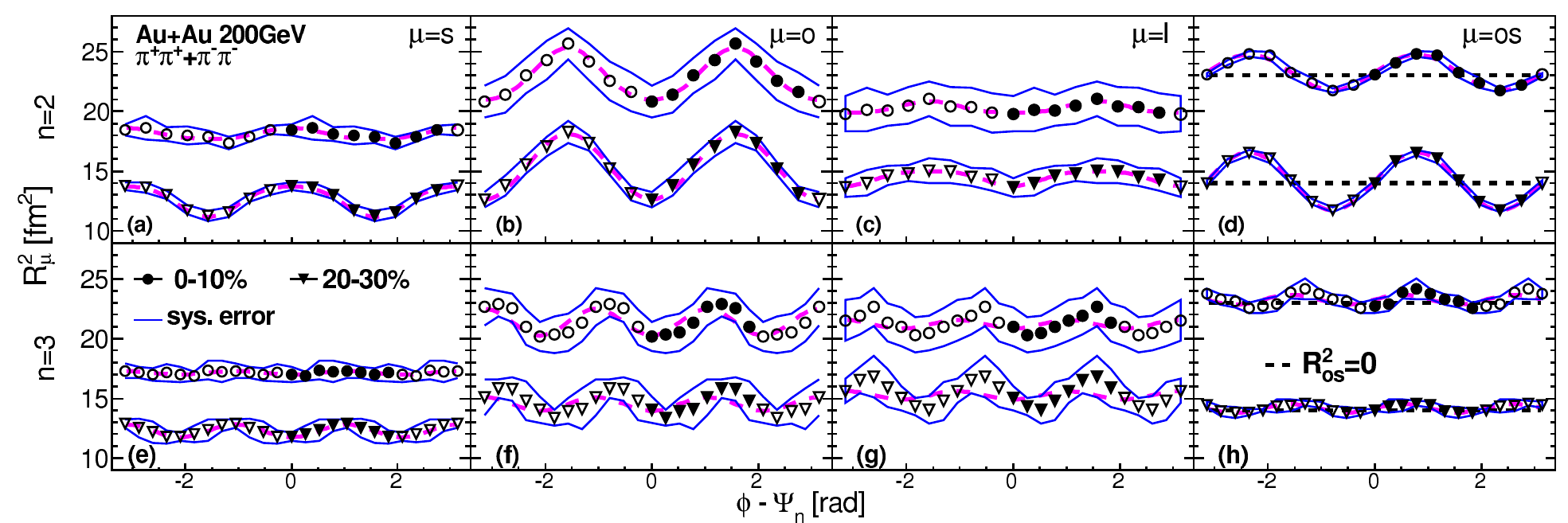}
\end{center}
\vspace{-0.5cm}
\caption{(Color online) The azimuthal dependence of $R_{s}^{2}$, 
$R_{o}^{2}$, $R_{l}^{2}$, and $R_{os}^{2}$ for charged pions in 
0.2\,$<$\,$k_{T}$\,$<$\,2.0 GeV/$c$ with respect to 2$^{\rm nd}$ 
(a-d) and 3$^{\rm rd}$-order (e-h) event plane in Au$+$Au collisions 
at \sqsn~=~200~GeV.  The $R_{os}^{2}$ is plotted relative to dashed 
lines representing $R_{os}^{2}$\,=\,0. 
The filled symbols
show the extracted HBT radii and the open symbols are reflected by
symmetry around $\phi-\Psi_{n}$\,=\,0. 
Bands of two thin lines show the systematic uncertainties and 
dashed lines show the fit lines by Eq.~(\ref{cosine_series}).}
\label{fig1}
\vspace{-0.1cm}
\end{figure*}

The experimentally measured correlation function is defined as 
$A(q)/B(q)$, where $A(q)$ is the relative-momentum 
distribution of all combinations of identified pion pairs in the 
same event, and $B(q)$ is the event-mixed background distribution 
of pairs formed from pions from different events, but with similar 
event centralities, vertex positions, and 2$^{\rm nd}$ 
(3$^{\rm rd}$-order) event planes.  To remove ghost tracks and detector 
inefficiencies, pairs with either $\Delta z$\,$<$\,5 cm and 
$\Delta\phi$\,$<$\,0.07 or $\Delta z$\,$<$\,70 cm and 
$\Delta\phi$\,$<$\,0.02 at the drift chamber were removed from the 
analysis, as were tracks separated by less than 17 cm at the front 
face of the electromagnetic calorimeters. The correlation functions 
were also binned according to the centrality of the event and 
the momentum  
of the pion pair. Positive and negative pion pairs were combined to 
cancel charge-dependent acceptance effect~\cite{star_pionhbt}.

A three-dimensional analysis was performed with the Bertsch-Pratt 
parameterization assuming a Gaussian 
source~\cite{ScottPara,BertschPara}:
\begin{equation}
G = \exp( -R_{s}^{2}q_{s}^{2} -R_{o}^{2}q_{o}^{2} -R_{l}^{2}q_{l}^{2} -2R_{os}^{2}q_{s}q_{o} ).
\end{equation}
In this framework, the relative momentum ${\bf q}$ is decomposed 
into $q_{l}$, $q_{o}$ and $q_{s}$, where $q_{l}$ denotes the beam 
direction, $q_{o}$ is perpendicular to $q_{l}$ and parallel to the 
mean transverse momentum of the pair 
$\vec{k}_{T}$\,=\,$(\vec{p}_{1T}+\vec{p}_{2T})/2$, and $q_{s}$ is 
perpendicular to both $q_{l}$ and $q_{o}$. The $R_\mu$ 
($\mu$\,=\,s,o,l) Gaussian parameters provide information on the 
size of the emission region in each direction, but $R_{o}$ and 
(to a lesser extent) $R_{l}$ include contributions from the emission 
duration and all are influenced by position-momentum correlations.  
The $R_{os}$ is a cross term that arises from asymmetries in the 
emission region~\cite{femtoscopy}. The analysis was performed in the 
longitudinally co-moving system, where $p_{1z}$\,=\,$-p_{2z}$. The 
measured correlation functions were fit by:
\begin{equation}
C_{2} = N [ ( \lambda (1+G) ) F_{c} + (1-\lambda)],
\label{eq:sinyukov}
\end{equation}
where $N$ is a normalization factor and $F_{c}$ is the Coulomb 
correction factor evaluated using a Coulomb wave function. 
Equation~(\ref{eq:sinyukov}) is based on the core-halo 
model~\cite{Bowler, Sinyukov}, which divides the source into two 
regions: a central core that contributes to the quantum 
interference, and a long-range component that includes the decay of 
long-lived particles having a negligible Coulomb interaction 
and a quantum statistical interference that occurs in a relative 
momentum range that is too small to be resolved experimentally. 
The fraction of pairs in the core is given by $\lambda$.

Finite event-plane resolution reduces the oscillation amplitude of 
HBT radii relative to the event plane. In this analysis, a 
model-independent correction suggested in~\cite{bw_hbt} was applied 
to $A(q)$ and $B(q)$. The correction factor is 54\% (32\%) for the 
2$^{\rm nd}$-order (3$^{\rm rd}$-order) event planes in 0\%--10\% 
centrality.  As a crosscheck, the oscillation amplitude was also 
corrected by dividing by ${\rm Res}(\Psi_{n})$~\cite{CERES}. 
Both methods applied to the 2$^{\rm nd}$- and 3$^{\rm rd}$-order 
event-plane dependence are consistent within systematic 
uncertainties. The effect of momentum resolution was studied using 
{\sc geant} simulations following previous 
analyses~\cite{BE_NA49,star_pionhbt} and its impact is negligible on 
the extracted radii ($<$\,1\%).


Systematic uncertainties were estimated by the variation of single 
track cuts, pair selection cuts, and input source size for the 
Coulomb wave function.   Also incorporated were the variations when 
using alternate event 
plane definitions from the forward, backward, and combined RXNPs. 
Total systematic uncertainties for 
$R_{s}^{2}$ and $R_{o}^{2}$ are not more than 5\% (12\%) and 7\% 
(17\%) for the 2$^{\rm nd}$-order (3$^{\rm rd}$-order) event plane,
respectively.


Figure \ref{fig1} shows $R_{s}^{2}$, $R_{o}^{2}$, $R_{l}^{2}$, and 
$R_{os}^{2}$ for pions as functions of azimuthal angle $\phi$ with 
respect to $\Psi_{2}$ and $\Psi_{3}$ for two centrality bins, where 
$\langle k_{T} \rangle \approx 0.53$ GeV/$c$. The filled symbols 
show the extracted HBT radii and the open symbols are reflected by 
symmetry around $\phi-\Psi_{n}$\,=\,0.  For the 0\%--10\% bin, 
$R_{s}^{2}$ shows a very weak oscillation relative to both 
$\Psi_{2}$ and $\Psi_{3}$, while $R_{o}^{2}$ clearly exhibits a 
stronger oscillation.  For the 20\%--30\% bin, $R_{s}^{2}$ and 
$R_{o}^{2}$ for $\Psi_{2}$ show opposite-sign oscillations, as 
expected for an elliptical source viewed from in-plane and 
out-of-plane axes~\cite{star_azhbt}. For $\Psi_{3}$, $R^{2}_{s}$ 
shows a weaker angular dependence of the same sign as $R^{2}_{o}$.


The oscillation amplitudes were extracted by fitting the angular 
dependence of $R^{2}_{\mu}$ to the functional form:
\begin{eqnarray}
R_{\mu}^{2} \!\!&=&\!\! R_{\mu,0}^{2} + 2\!\!\!\! \sum_{n=m,2m} \!\!\!\!R_{\mu,n}^{2} \cos[ n (\phi-\Psi_{m}) ] \,\,\; (\mu=s, o, l), \nonumber \\
R_{\mu}^{2} \!\!&=&\!\! 2\!\!\!\!  \sum_{n=m,2m} \!\!\!\!R_{\mu,n}^{2} \sin[ n (\phi-\Psi_{m}) ] \,\,\; (\mu=os), 
\label{cosine_series}
\end{eqnarray}
where $R_{\mu,n}^{2}$ are the Fourier coefficients~\cite{R_mu}.

\begin{figure}[tb]
\begin{center}
\includegraphics[width=1.0\linewidth]{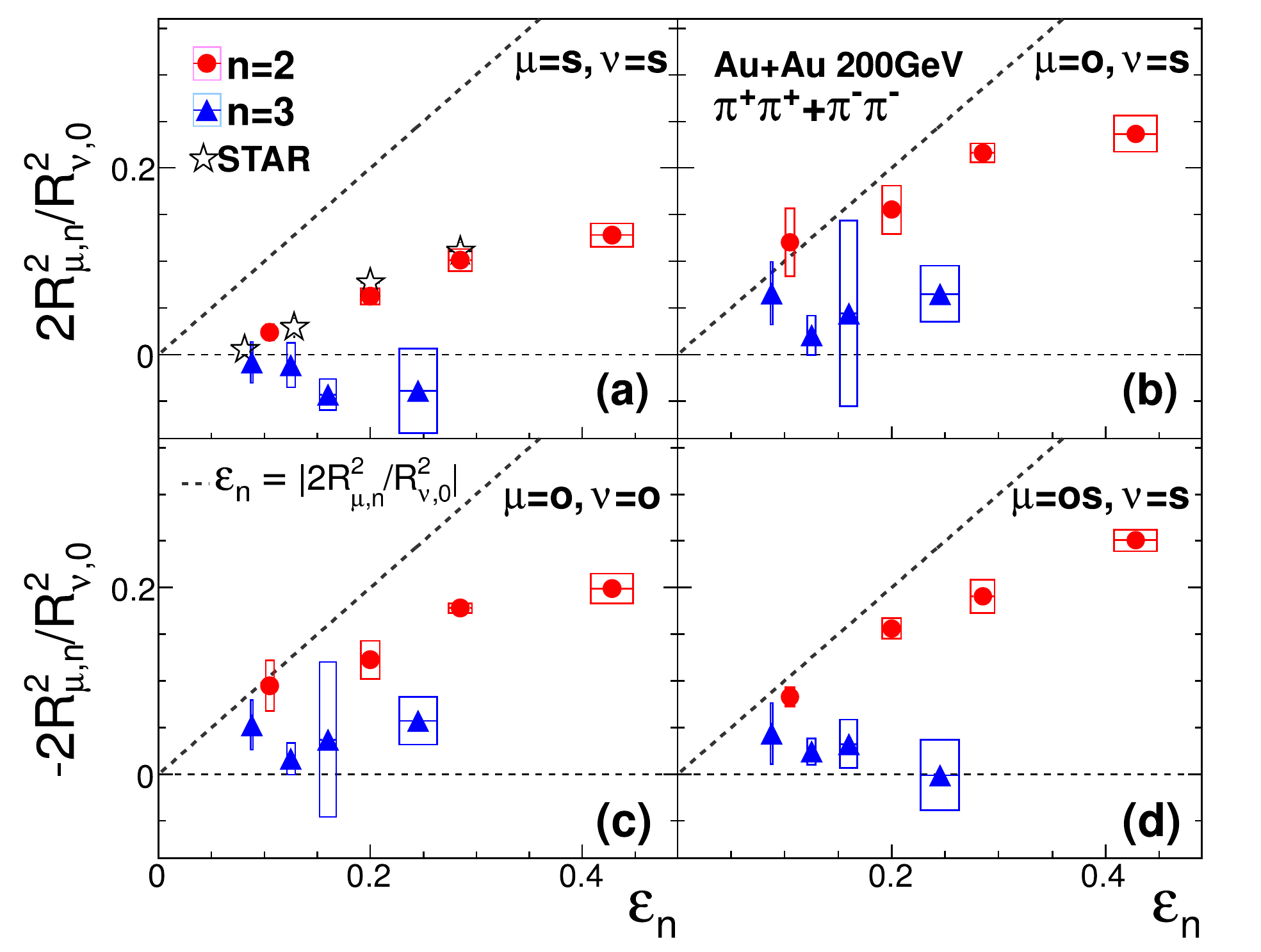}
\end{center}
\vspace{-0.5cm}
\caption{(Color online) The solid points are the oscillation 
amplitudes relative to the average of HBT radii as a function of 
initial spatial anisotropy ($\varepsilon_{n}$), which are calculated 
using the Glauber model.  Boxes show the systematic uncertainties.  
Open star symbols are the $\varepsilon_{final}$ from 
STAR~\cite{star_azhbt}.  Dashed lines indicate the line of 
$\varepsilon_{n}$=$|2R_{\mu,n}^{2}/R_{\nu,0}^{2}|$.}

\label{fig2}\vspace{-3pt}
\end{figure}

\begin{figure}[tbh]
\begin{center}
\includegraphics[width=1.0\linewidth]{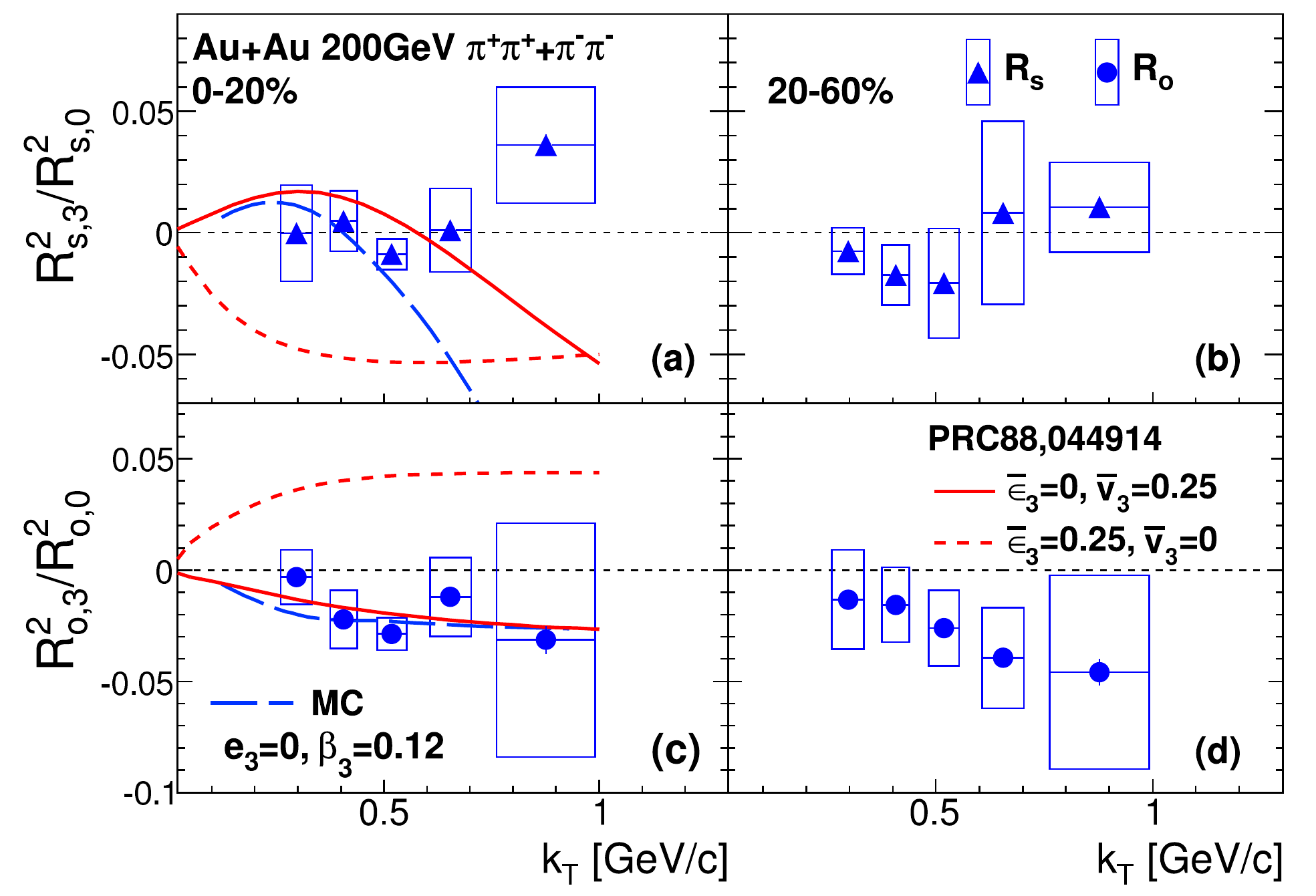}
\end{center}
\vspace{-0.5cm}
\caption{(Color online) $k_{T}$ dependence of $R_{s}^{2}$ ((a),(b)) 
and $R_{o}^{2}$ ((c),(d)) amplitudes relative to their averages for 
3$^{\rm rd}$-order event plane in two centrality bins. Calculations 
of the Gaussian source model~\cite{UliPaper} are shown as solid and 
short-dashed (red) curves, where the values are scaled by 0.3.  
Calculations using the MC-simulation are shown as long-dashed (blue) 
curves.}
\label{fig3}
\end{figure}

Figure~\ref{fig2} shows the amplitudes relative to the average of 
$R_{s}^{2}$, $R_{o}^{2}$, and $R_{os}^{2}$, 
$2R^{2}_{\mu,n}/R^{2}_{\nu,0}$, as functions of initial eccentricity 
($\varepsilon_{2}$) and triangularity ($\varepsilon_{3}$). Each 
$\varepsilon_{n}$ is calculated by Monte-Carlo Glauber simulation as 
given in Ref.~\cite{glauber,glauber_en} and decreases with 
increasing centrality, however the centrality dependence of 
$\varepsilon_{3}$ is weaker than that of $\varepsilon_{2}$.

The $2R_{s,2}^{2}/R_{s,0}^{2}$ (Fig.\,\ref{fig2}(a)) is sensitive to 
the final source eccentricity ($\varepsilon_{\rm final}$) at 
freeze-out~\cite{bw_hbt}, and approaches the whole source 
eccentricity in the limit of $k_{T}$\,=\,$0$.\,\,\,Our results for 
the $\Psi_{2}$ dependence are consistent with the STAR 
experiment~\cite{star_azhbt}. We note that the $\varepsilon_{\rm 
final}$ defined from $R_{s}$ has a systematic uncertainty of 30\% 
due to the assumption of space-momentum correlation in the 
Blast-Wave model~\cite{bw_hbt}. The positive value of 
$\varepsilon_{\rm final}$ indicates that the source shape still 
retains the initial shape extended out-of-plane, though reduced in 
magnitude.
%
Other combinations of $|2R^2_{\mu,2}/R^2_{\nu,0}|$ also have similar 
$\varepsilon_{n}$ dependence, but are larger than 
$2R^{2}_{s,2}/R^{2}_{s,0}$. They include contributions from the 
emission duration and will have different sensitivity to the 
dynamics~\cite{hydro_asHBT}.
%
The $2R_{s,3}^{2}/R_{s,0}^{2}$ are less than or equal to zero, which 
seems to be an opposite trend to other combinations, as noted 
already in Fig.~\ref{fig1}.  For all amplitudes, the values for 
3$^{\rm rd}$-order are small compared to those for 2$^{\rm 
nd}$-order.


It is well known that the HBT radii are influenced by the presence 
of dynamical correlations between momentum and spatial distributions 
at the time of freeze-out~\cite{Hama,Akkelin}, as evident in the 
transverse pair momentum $k_{T}$ dependence of the radii.  
Figure~\ref{fig3} shows these results for the 3$^{\rm rd}$-order 
oscillation amplitudes.  The $R_{o,3}^{2}/R_{o,0}^{2}$ decreases 
with $k_{T}$, whereas $R_{s,3}^{2}/R_{s,0}^{2}$ does not show a 
significant dependence.


Although the reduced 3$^{\rm rd}$-order anisotropy in 
Fig.~\ref{fig3} may indicate small triangular deformation at 
freeze-out, its interpretation is complicated by the influence of 
dynamical correlations from the triangular flow~\cite{UliPaper}.  
To illustrate the different contributions of these effects we show 
separately the $k_T$ dependence for a source with radial symmetry 
and triangular flow ($\bar{\epsilon}_3$=0, $\bar{v}_{3}$=0.25) and a 
source with triangular deformation and radial flow 
($\bar{\epsilon}_3$=0.25, $\bar{v}_{3}$=0)~\cite{GSmodel}.  The 
model curves are taken from~\cite{UliPaper}, but the radii are 
scaled by 0.3 to fit within the range of the data.  The $R^2_{o,3}$ 
favors the deformed flow scenario, while the $R^2_{s,3}$ matches the 
deformed flow only at lower $k_{T}$.


To disentangle the relative contributions of spatial and 
flow anisotropy to the azimuthal dependence of HBT radii, we have 
performed a Monte-Carlo simulation introducing the spatial 
anisotropy and collective flow with anisotropic modulation at 
freeze-out. The assumptions of this model are similar to those 
adopted in the Blast-Wave (BW) model~\cite{bw_hbt, Spectra_bw}, 
generalized for 3$^{\rm rd}$-order modulation, and do not include 
effects such as viscosity and source opacity. The particle 
distributions in the transverse plane were parameterized with a 
Woods-Saxon function, $\Omega(r)$\,=\,$1/(1+\exp[(r-R)/a])$. To 
control the final source triangularity, we introduced a parameter 
$e_{3}$ into the radius parameter R in $\Omega(r)$ as follows:
\begin{eqnarray}
R &=& R_{0} \, (1-2e_{3} \cos[3(\phi-\Phi)]), \label{eq_R} \\
\beta_{T} &=& \beta_{0} \, ( 1 + 2 \beta_{3} \cos[3(\phi-\Phi)] ), 
\label{eq_Beta}
\end{eqnarray}
where $\phi$ is the azimuthal angle of particle positions, $\Phi$ is 
reference angle of the spatial anisotropy and triangular flow, and 
$R_{0}$ is average radius. To take the collective flow into account, 
generated particles were boosted in the transverse radial direction 
with a velocity $\beta_{T}$ in addition to their thermal velocities. 
We used a similar definition to the BW model~\cite{bw_hbt, 
Spectra_bw} as the flow rapidity 
$\rho(r)$\,=\,$(r/R)\tanh^{-1}(\beta_{T})$. In Eq.\,(\ref{eq_Beta}), 
$\beta_{0}$ represents the average of radial flow and $\beta_{3}$ is 
used to control the flow anisotropy. We assume that the particles 
are emitted with a Gaussian time distribution with $\Delta\tau$ 
standard deviation, which affects $R_{o}$, but not $R_{s}$. The 
effect of HBT interference was calculated by $\cos(\Delta {\bf x} 
\cdot {\bf q})$, where $\Delta {\bf x}$ and ${\bf q}$ are 4-vectors 
for relative distance and relative momentum of the pair. All other 
parameters except $e_{3}$ and $\beta_{3}$ were tuned to reproduce 
the strength of radial flow measured by $m_{T}$ 
spectra~\cite{Spectra_phenix} and the averages of HBT radii shown in 
Fig.~\ref{fig1}. For this analysis $\Delta\tau$ was set to 3.5 
fm/$c$ (2.7 fm/$c$) for 0\%--10\% (20\%--30\%) to achieve better 
agreement with the average of $R_{o}^{2}$. A simulation result with 
$e_{3}$\,=\,0 and $\beta_{3}$\,=\,0.12 is shown in Fig.~\ref{fig3}, 
displaying a trend that is qualitatively consistent with 
Ref.~\cite{UliPaper}.


\begin{figure}[tbh]
\begin{center}
\includegraphics[width=1.0\linewidth]{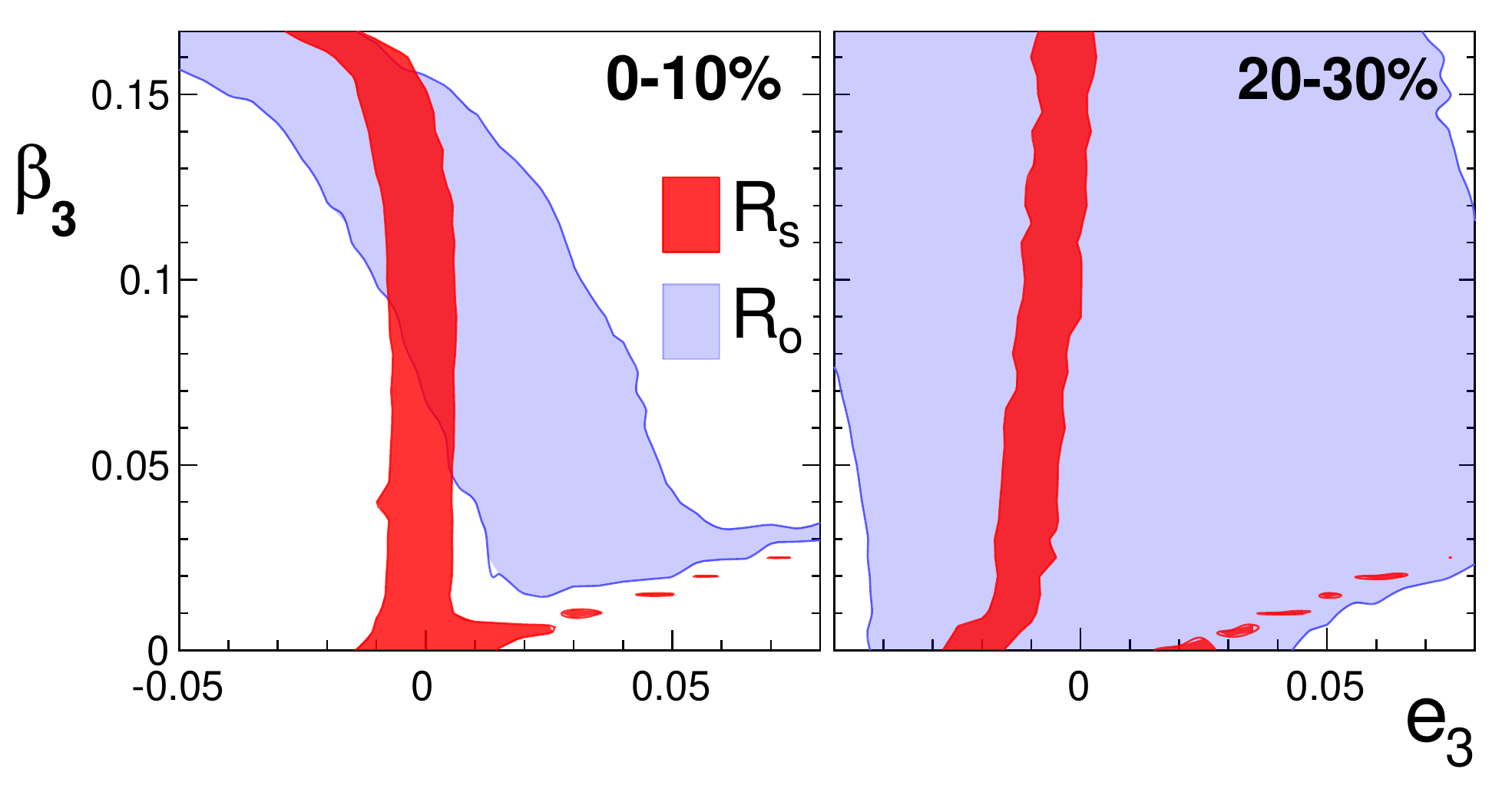}
\end{center}
\vspace{-0.5cm}
\caption{(Color online) $\chi^{2}$ contours representing the 
difference between data and simulation in 
$2R_{\mu,2}^{2}/R_{\mu,0}^{2}$ ($\mu$=$s$, $o$), as functions of 
$e_{3}$ and $\beta_{3}$. Shaded areas represent $\chi^2$ less than 
unity and constrained by the experimental uncertainty.}
\label{fig4}\vspace{-4pt}
\end{figure}

To understand how the data may constrain these values, we 
have performed a least-square fit for $e_{3}$ and $\beta_{3}$. 
Figure~\ref{fig4} shows the contour plots of $\chi^{2}$ defined by 
$(( [2R^{2}_{\mu,3}/R^{2}_{\mu,0}]^{\rm exp} - 
[2R^{2}_{\mu,3}/R^{2}_{\mu,0}]^{\rm sim}) / E )^{2}$, where $E$ is 
the experimental uncertainty. The value of $e_{3}$ is well 
constrained by the measured value of $R_{s}^{2}$, and indicates that 
the final triangularity is very close to zero.  The inclusion of 
$R_{o}^{2}$ favors a positive value for $\beta_{3}$ for 0\%--10\%, but 
does not add much information to 20\%--30\%, where a slightly negative 
value of $e_{3}$ is favored by $R_{s}^{2}$.  We note that the 
discrepancy at high $k_{T}$ remains, but the data integrated over 
$k_{T}$ is primarily influenced by lower $k_{T}$ pairs.
Detailed comparison with a realistic hydrodynamic model 
(e.g.,~\cite{UliPaper,bozek}) will be a key to fully understand 
the results.

%

In summary, we have presented results on the azimuthal dependence of 
charged-pion HBT radii with respect to 2$^{\rm nd}$- and 
3$^{\rm rd}$-order event planes in Au$+$Au collisions at 
\sqsn~=~200~GeV.  The results for the 2$^{\rm nd}$-order event plane 
dependence indicate that in noncentral events the source starts with 
an initial elliptical distribution and ends with an elliptical 
distribution at freeze-out, but with a diluted eccentricity due to 
the medium expansion.  For the 3$^{\rm rd}$-order event plane 
results, the observed $R_{o}^{2}$ oscillation may come from flow 
anisotropy, but the small $R_{s}^{2}$ oscillation with the same sign 
as $R_{o}^{2}$ in noncentral collisions may imply that the source 
expansion with triangular flow inverts the initial triangular shape. 
A Monte-Carlo simulation for an expanding triangular transverse 
distribution produced results consistent with this interpretation. 
Comparisons with an event-by-event hydrodynamic model will be needed 
to reveal the relation of spatial and hydrodynamical flow anisotropy 
at freeze-out, as well as to provide further constraints on the 
hydrodynamic evolution in relativistic heavy ion collisions.




We thank the staff of the Collider-Accelerator and Physics
Departments at Brookhaven National Laboratory and the staff of
the other PHENIX participating institutions for their vital
contributions.  We acknowledge support from the 
Office of Nuclear Physics in the
Office of Science of the Department of Energy, 
the National Science Foundation, 
Abilene Christian University Research Council, 
Research Foundation of SUNY, and 
Dean of the College of Arts and Sciences, Vanderbilt University (U.S.A),
Ministry of Education, Culture, Sports, Science, and Technology
and the Japan Society for the Promotion of Science (Japan),
Conselho Nacional de Desenvolvimento Cient\'{\i}fico e
Tecnol{\'o}gico and Funda\c c{\~a}o de Amparo {\`a} Pesquisa do
Estado de S{\~a}o Paulo (Brazil),
Natural Science Foundation of China (P.~R.~China),
Ministry of Education, Youth and Sports (Czech Republic),
Centre National de la Recherche Scientifique, Commissariat
{\`a} l'{\'E}nergie Atomique, and Institut National de Physique
Nucl{\'e}aire et de Physique des Particules (France),
Bundesministerium f\"ur Bildung und Forschung, Deutscher
Akademischer Austausch Dienst, and Alexander von Humboldt Stiftung (Germany),
Hungarian National Science Fund, OTKA (Hungary), 
Department of Atomic Energy and Department of Science and Technology (India), 
Israel Science Foundation (Israel), 
National Research Foundation and WCU program of the 
Ministry Education Science and Technology (Korea),
Physics Department, Lahore University of Management Sciences (Pakistan),
Ministry of Education and Science, Russian Academy of Sciences,
Federal Agency of Atomic Energy (Russia),
VR and Wallenberg Foundation (Sweden), 
the U.S. Civilian Research and Development Foundation for the
Independent States of the Former Soviet Union, 
the US-Hungarian Fulbright Foundation for Educational Exchange,
and the US-Israel Binational Science Foundation.



\begin{thebibliography}{40}
\bibitem[{R_m()}]{R_mu}
\bibinfo{note}{The $R_{\mu,n}^{2}$ as a fitting parameter could be negative,
  which means a different phase of the cosine function.}
\end{thebibliography}

\end{document}